\documentclass[12pt]{article}
\usepackage{amsmath}
\usepackage{graphicx,psfrag,epsf}
\usepackage{enumerate}
\usepackage{enumitem}
\usepackage{natbib}
\usepackage{url} % not crucial - just used below for the URL

\usepackage{indentfirst}
\usepackage{subcaption}
\usepackage[font={small,it, sf}]{caption}
\usepackage{float}
\usepackage{amssymb, amsthm, bbm, mathtools, mathabx}
\usepackage{hyperref}
\usepackage{cleveref}
\usepackage{xr}
\usepackage[table,xcdraw]{xcolor}

\usepackage{ulem}

\newtheorem{theorem}{Theorem}
\newtheorem*{theorem*}{Theorem}

\theoremstyle{definition}

\theoremstyle{definition}

\DeclareUnicodeCharacter{03B2}{$\beta$}

\newcommand{\blind}{1}

\begin{document}

\def\spacingset#1{\renewcommand{\baselinestretch}%
{#1}\small\normalsize} \spacingset{1}

%%%%%%%%%%%%%%%%%%%%%%%%%%%%%%%%%%%%%%%%%%%%%%%%%%%%%%%%%%%%%%%%%%%%%%%%%%%%%%

\if1\blind
{
  \title{\bf Modeling Cell Developmental Trajectory using Multinomial Unbalanced Optimal Transport}
  \author{Junhao Zhu, Kevin Zhang \\
    Department of Statistical Sciences, University of Toronto,\\
    Zhaolei Zhang \\
    Department of Molecular Genetics, \\
    Donnelly Centre for Cellular and Biomolecular Research, \\
    University of Toronto, \\
    and \\
    Dehan Kong\\
    Department of Statistical Sciences, University of Toronto   \thanks{
    The first two authors contribute equally to the paper. The last two authors are joint corresponding authors. }\hspace{.2cm}\\}
  \maketitle
} \fi

\if0\blind
{
  \bigskip
  \bigskip
  \bigskip
  \begin{center}
    {\LARGE\bf Modeling Cell  Developmental Trajectory using Multinomial Unbalanced Optimal Transport}
\end{center}
  \medskip
} \fi

\bigskip
\begin{abstract}
Inferring developmental trajectories from snapshot single-cell RNA sequencing (scRNA-seq) data remains a fundamental challenge due to the destructive nature of sequencing and the high level of noise in cellular measurements. While optimal transport (OT) methods applied at the single-cell level have gained increasing attention, they often suffer from high estimation variance and substantial computational cost.
 We propose a metacell based Multinomial Unbalanced OT framework to reconstruct cellular developmental dynamics.   By aggregating similar cells into metacells prior to transport estimation, the proposed approach introduces structural constraints that improve both statistical stability and computational efficiency. Through analysis of a large-scale  mouse developmental atlas data, we demonstrate that metacell-level OT  provides more reliable estimates of transition probabilities and population growth rates than single-cell OT methods,  which can be sensitive to technical variation and computational artifacts. Our results accurately recover cell-type transitions congruent with biological ground truth, suggesting that metacell aggregation is not merely a computational convenience but a statistical necessity for reliable OT trajectory inference. We further establish non-asymptotic and asymptotic convergence guarantees, as well as bootstrap consistency for valid confidence interval construction, under a finite-mixture model with data-driven transport costs.
\end{abstract}

\noindent%
\vfill

\newpage
\spacingset{1.9} % DON'T change the spacing!
\section{Introduction}
\label{sec:intro}
Cellular development is a fundamentally dynamic process characterized by transitions between distinct biological states. These transitions are governed by an underlying genetic program, where the cell status at any given stage is  reflected in its gene expression profiles. Scientists are interested in understanding   cellular development, such as  how the transition from a zygote or a stem cell to a specific cell type or a specialized tissue,  particularly in the context of regenerative medicine \citep{keller2005embryonic}, embryo development and cellular reprogramming \citep{mothe2012advances}. The single-cell RNA sequencing (scRNA-seq) has emerged as an advanced technology to study the cellular development, allowing researchers to measure  the transcriptomes of  individual cells \citep{tanay2017scaling}. By providing a detailed profile of cellular heterogeneity, scRNA-seq enables the identification of different  cell types and even more subtle intermediate states based on their RNA expression profiles that were previously masked by  bulk sequencing. However,  scRNA-seq is intrinsically limited by its desturctive sampling design,  since  the sequencing process is destructive and it is difficult to obtain continuous longitudinal observation of the same individual cell over time. 

To overcome the snapshot nature of scRNA-seq data, researchers develop computational models to reconstruct the cellular developmental trajectories. These models treat cells as dynamic processes evolving in a high dimensional state  space and aim to infer both their developmental histories and future fates  from the data.  The mathematical models of  trajectory inference for scRNA-seq snapshot data mainly fall into three primary frameworks:  pseudo-time,  RNA velocity, and   optimal transport. While these methods all aim to recover temporal dynamics, they utilize different data structures and mathematical principles. The pseudo-time methods such as Monocle \citep{cao2019single} and Slingshot \citep{street2018slingshot} order cells along a latent one-dimensional  manifold based on their transcriptional similarity and prior knowledge. 
RNA-velocity models, such as scVelo \citep{bergen2020generalizing} and DeepVelo \citep{chen2022deepvelo}, characterize cells by leveraging splicing kinetics and differential equations to predict their immediate future states, thereby providing a velocity field that describes cellular state transitions. Both the pseudo-time and RNA-velocity models are in principle designed for a static population of cells collected at a single time point containing individual cells at various stages of a shared process. In contrast,    the OT framework, such as Waddington OT \citep{schiebinger2019optimal} and   more advanced methods  DestOT \citep{halmos2025dest} or     Moscot \citep{klein2025mapping}, mainly focuses on studies where samples are collected at  multiple developmental stage,   aims to model the temporal processes at a distribution level  and  links multiple snapshot of cellular population across  different collective  time points.
This evolving distribution perspective is deeply rooted in Waddington’s landscape \citep{waddington1957strategy}  and optimal transport (OT) \citep{monge1781memoire}. Waddington’s landscape conceptualizes cellular development as a series of rocks rolling downhill, where  cells will fall into specific valleys that correspond to distinct developmental outcomes or stable cell states. OT,  originally proposed by \citet{monge1781memoire}, aims  to model the efficient transition between distributions, serves as the mathematical bridge for reconstructing the developmental paths. As demonstrated by \citet{schiebinger2019optimal}, by assuming that cellular populations transit in the ``least action'' or the  most efficient  way, OT can identify the probable trajectories between experimental snapshots. In this manuscript, we mainly focus on the OT setting that data are  collected at  multiple developmental stage. 

Despite the prominence of OT in trajectory inference, its application at the individual-cell granularity introduces significant statistical and computational challenges. First, while established frameworks such as Waddington-OT and Moscot typically rely on an optimal transport  principle, they assume access to an uncontaminated measurement of the underlying distribution. However, scRNA-seq data are inherently corrupted by measurement noise and technical variation \citep{sarkar2021separating}. Direct application of OT to such noisy   data lacks rigorous statistical guarantees, as the resulting transport maps are contaminated by  technical noise. Secondly, individual-level inference has been shown to induce substantial  variance, particularly concerning cellular growth rates within the same cell type \citep{halmos2025dest}. The estimated differentiation also lacks a corresponding method for uncertainty quantification. 
Finally, the computational complexity of solving OT on large-scale scRNA-seq dataset remains a significant bottleneck. 

To address these challenges, we introduce Trajectory Inference with Multinomial Optimal Transport (TIMO), a framework that shifts the modeling focus from individual cells to trajectories of metacells. TIMO contains two steps. First, we aggregate
homogeneous group of cells into metacells, treating differences among cells within each metacell as technical noise rather than biological variation \citep{baran2019metacell, persad2023seacells, sarkar2021separating}.  
Second, we solve a semi-relaxed optimal transport problem under the constraint that all cells within the same metacell share identical transport probabilities, reflecting their interpretation as a common biological state. This constrained formulation is equivalent to solving a multinomial optimal transport problem with marginal distributions defined by sizes of metacells and costs defined by squared distances between metacell centroids. To evaluate the   performance of TIMO, we apply it to a large atlas dataset of early mouse development containing expression profiles of 1.7 million cells measured across 20 developmental stages from embryonic day E3.5 to E13.5. TIMO produces biologically  reliable estimates of cell-type transition probabilities and population growth rates, as supported by agreement with curated developmental transitions, improved growth-rate distortion metrics, and a larger number of identified growth-related genes. In addition, performing OT at the metacell level substantially reduces computational cost.
Finally, we study the non-asymptotic and the asymptotic  properties of our proposed estimators, as well as the bootstrap consistency for valid confidence interval construction. To the best of our knowledge, this is the first work establishing statistical theories for multinomial unbalanced optimal transport with random transport costs, where the cost matrix is estimated from the data through metacell centroids.

We structure the rest of the paper,  as follows. In Section \ref{sec:datades}, we introduce background of single-cell developmental data. We then introduce the concept of metacell,  optimal transport and our method TIMO in Section \ref{sec:method}.   Application of  TIMO to a large-scale atlas data of mouse early embryo is presented in Section \ref{sec:data}. Section \ref{sec:theory} contains theoretical   properties of our proposed estimators.  In Section \ref{sec:diss}, 
 we end with a brief discussion.

\begin{figure}[]
\centering  
\includegraphics[width=0.95\textwidth]{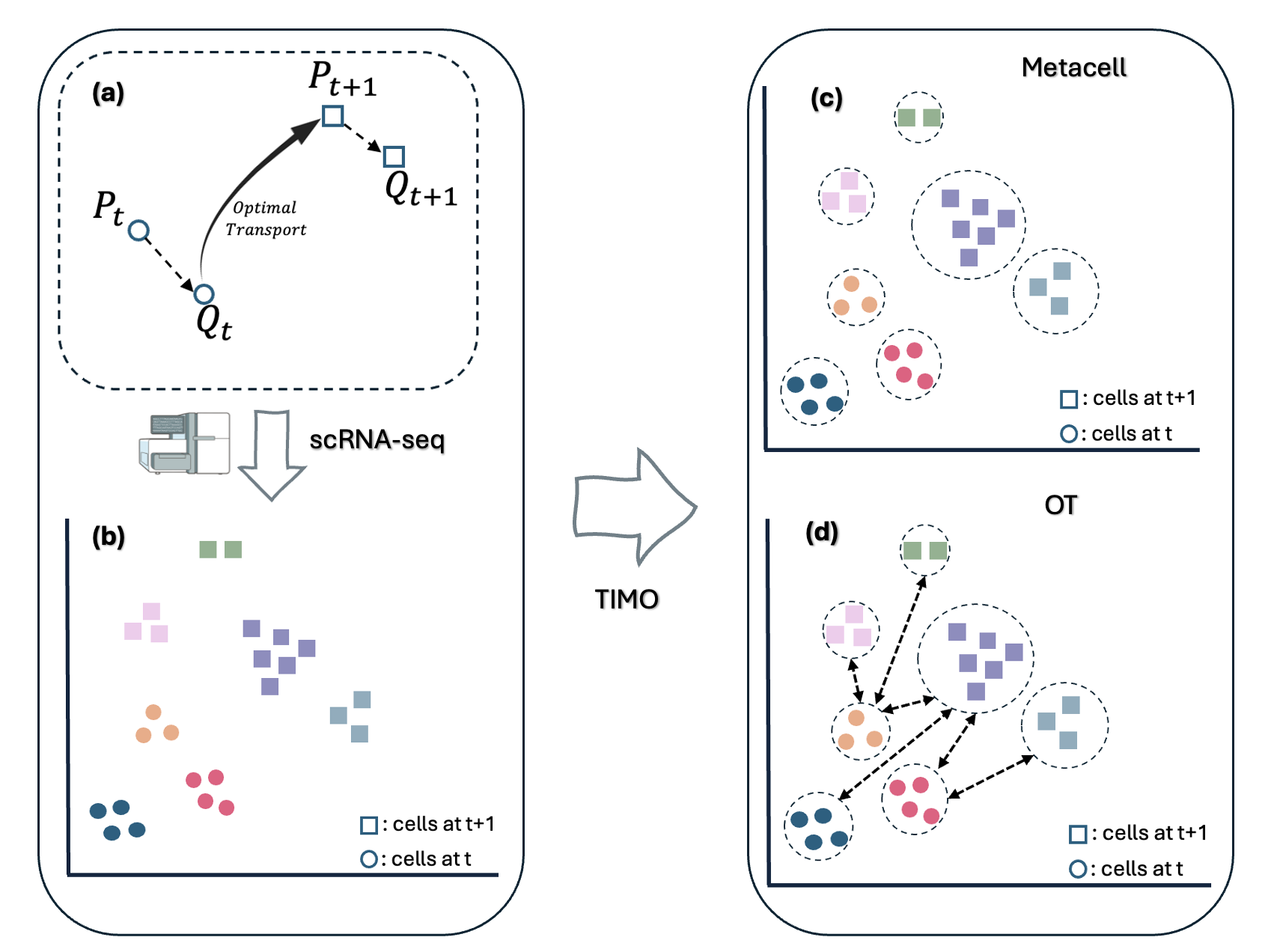}  
\caption{\textbf{Schematic Overview of TIMO to two consecutive time points $t$ and $t+1$.}\  \textbf{(a):} Conceptual assumption for scRNA-seq temporal data. $P_t$ and $P_{t+1}$ are  the probability distributions of the cellular biological states, while $Q_t$ and $Q_{t+1}$  are the after-growth distributions. $P_t$ first undergoes growth to $Q_t$ then transfers to $P_{t+1}$ following the optimal transport principle.  \ \textbf{(b):} The scRNA-seq   data at time $t$  and $t+1$. Each circle represents a cell at $t$ while each cube represents a cell at $t+1$. \ \textbf{(c):}   The first step of TIMO is to group cells into metacells. Within each metacell the difference is considered to be technical variation. \ \textbf{(d):}  Solving the semi-relaxed optimal transport problem to transport metacells across time $t$ and time $t+1$.     } 
\label{Fig.main}
\end{figure}

% \section{Method}
% \label{sec:method}
% In this section, we introduce the background of our method, including the background of the scRNA-seq data in the developmental context, a brief review of optimal transport and the concept of metacell.  After laying out the required preliminary,  we illustrate the motivation and detail our proposed method called TIMO   built upon the  metacell construction and the optimal transport principle. 

\section{Single-cell developmental data}\label{sec:datades}
We begin by introducing the background of the scRNA-seq data in the developmental context. During mammalian development, a single zygote rapidly multiplies and differentiates into millions of cells of different biological states. This process generates the vast array of molecular programs and diverse cell types required to form the complicated organism. A fundamental challenge in this field is to reconstruct cellular trajectories and map the transformation of cells into the diverse cellular landscape. Addressing this challenge requires us answering two  questions \citep{schiebinger2019optimal,qiu2022systematic}: 1. what are the biological states of cells at different developmental stages? 
Given these states, what are the probable ancestral origins and future fates of these cells? The first question can be answered by the  single-cell RNA sequencing \citep{tanay2017scaling}. Typically, at each time point or developmental stage,  the scRNA-seq provides whole-transcriptome gene expression profiles for individual cells within a  population, where these profiles serve as a high-dimensional representation of cellular states.  However, since  the  scRNA-seq is a destructive measurement process, gene expression levels of cells cannot be observed at later time points after they have been measured. Ideally we want to observe the   cellular developmental trajectory. 
Instead,  we only observe several data matrices of cells' expression profiles, or equivalently an empirical distribution of gene expression profiles within a cellular population  like a snapshot at each  developmental stage of interest. Suppose the biological state of each individual cell is represented by a multivariate vector in $\mathbb{R}^D$, where $D
$ may correspond to the number of genes measured in scRNA-seq, the number of detected chromatin accessibility peaks in single-cell ATAC-seq, or more generally the dimension of features obtained through dimensionality reduction such as Seurat \citep{butler2018integrating} in general single-cell omics data. A developmental process can then be modeled as a collection of time-varying distributions $\{P_t: t=1,\ldots,T\}$ defined over this $D$-dimensional state space. In practice, cells are typically sequenced only at a finite set of time points determined by the experimental design, and therefore the time index $t$ can be treated as a discrete set of positive integers. 
At each time point $t$, due to the snapshot nature of scRNA-seq experiments and the presence of technical noise, we observe mutually independent samples $\{X_{t,i}: t=1,\ldots,T,\; i=1,\ldots,N_t\}$, where in our case there are $\sum_t N_t\approx 1.7\times 10^6$ cells being sequenced.   Specifically,  (i) $X_{t_1,i_1}$ is independent with $X_{t_2,i_2}$
whenever $t_1\neq t_2 $  or $i_1\neq i_2$, and (ii) each observation follows  the data generated scheme  $X_{t,i} = \mu_{t} +\epsilon$, where $\mu_t\sim P_t$ and $\epsilon$ is the random noise,   reflecting technical noise contamination in the measurements.  

In our case study, we analyze the mouse embryogenesis atlas single-cell dataset compiled by \cite{qiu2022systematic}. This dataset consists of snapshot measurements collected across $T=20$ time points, spanning embryonic days E3.5 to E13.5 and covering three major stages: pre-gastrulation (E3.5–6.5), gastrulation (E6.5–8.5), and organogenesis (E8.5–E13.5). The atlas contains gene expression profiles for approximately 1.7 million cells across more than 90 cell types and is designed to investigate how the inner cell mass (ICM) at E3.5, also known as the embryoblast within the mammalian blastocyst, gives rise to diverse cell types and subtypes throughout development.

Due to this snapshot nature,  realizations from the coupling or joint distribution $\mathbb{P}_{t,t+1}$ for any $t$ are never directly observable. However,  the quantities of primary interest are the ancestor distributions and descendant distributions  that are defined by the joint distribution and characterize cellular dynamics across time. 
Specifically, suppose a cell with observed state $X \in \mathbb{R}^D$ is observed at time $t$ with $X \sim P_t$. The ancestor distribution describes the conditional distribution of its possible origins at the previous time point $t-1$, while the descendant distribution characterizes the conditional distribution of its potential future states at time $t+1$. Mathematically, these correspond to the conditional distributions induced by the joint distribution ${P}_{t-1,t}$ and ${P}_{t,t+1}$.
Without any additional assumptions or structural constraints, the coupling cannot be uniquely recovered from the independent   marginal distributions, since infinitely many joint distributions can generate the same time-varying marginal distributions. To make estimation feasible, we follow prior work \citep{schiebinger2019optimal,shi2022energy} and assume that cells evolve between two consecutive time points according to a least-effort principle in the space of biological states. This assumption is also known as the optimal transport principle, and  introduces additional structure that enables estimation of the joint distribution from the observed marginals. We will then introduce the concept of optimal transport in Section \ref{sec:method}. 
% Moreover, we assume that    the scRNA-seq measurement follows a biological signal plus technical noise generating process \citep{sarkar2021separating} and utilize the recently  developed metacell methods \citep{baran2019metacell,persad2023seacells} to address the second problem and thus statistical guarantees such as non-asymptotic sample bound and asymptotic distribution can be achieved under mild conditions. 

\section{Method}
\label{sec:method}
In this section, we introduce the background of our method, including  a brief review of optimal transport and the concept of metacell.  After laying out the required preliminary,  we illustrate the motivation and detail our proposed method called TIMO   built upon the  metacell construction and the optimal transport principle. 
\subsection{Optimal transport}
Optimal transport was first introduced by \citet{monge1781memoire} and formalized by \citet{kantorovich1942translocation} to model the transfer process between two distributions. The rationale of  optimal transport  is that it assumes the population evolves in a cost-efficient fashion. For any two probability distributions  $P$ and $Q$ over the Euclidean space $\mathbb{R}^D$, $ \Pi({P}, {Q})$ denotes the set of couplings or joint distributions with marginal distributions ${P}$ and ${Q}$ throughout this manuscript. Then the optimal transport aims to solve the following optimization problem: 
\begin{equation}\label{ot0}
    \pi_{P, {Q}}=\arg\min_{\pi \in \Pi({P},  {Q})}
\int_{ \mathbb{R}^D \times  \mathbb{R}^D}
c(x,y)\, d\pi(x,y),
\end{equation}
where ${\pi_{{P},{Q}}}$ is   the optimal coupling between ${P}$ and ${Q}$. When dealing with finite samples, the optimal coupling ${\pi_{{P},{Q}}}$ will be a matrix representing the joint probability and is often called the optimal transport matrix. 
In practice, the squared Euclidean cost $c(x,y)=\|x-y\|_2^2$ is commonly used due to its favorable theoretical properties \citep{villani2009optimal,peyre2019computational,chewi2025statistical}.
A direct application of OT to the scRNA-seq snapshot data is to solve the optimization problem in \eqref{ot0} with empirical distributions derived from  two consecutive time points. 
\begin{equation}\label{vanillaot}
\begin{aligned}
\hat{\boldsymbol{\pi}}
= &\arg\min_{\boldsymbol{\pi} \in \mathbb{R}_+^{N_t\times N_{t+1}}}
\Bigg\{
 \sum_{i=1}^{N_t}\sum_{j=1}^{N_{t+1}} \pi_{ij}\, \|X_{t,i}-X_{t+1,j}\|_2^2 
\Bigg\},
\\ &\text{s.t.} \sum_{i=1}^{N_{t}} \pi_{ij} = \frac{1}{N_{t+1}}, ~~\sum_{j=1}^{N_{t+1}} \pi_{ij} = \frac{1}{N_{t}},~~\pi_{ij}\geq 0. 
% \\
% & \text{where} \qquad r_i := \sum_{j=1}^{N_{t+1}}\Pi_{ij} \qquad\text{and}\qquad
% c_j := \sum_{i=1}^{N_t}\Pi_{ij}. \qquad
\end{aligned}
\end{equation}
In this case, $\boldsymbol{\hat\pi}$ is  an $N_t\times N_{t+1}$ matrix, where the $(i,j)$-entry $\pi_{ij}$ represents the probability that a randomly sampled cell has biological states closed to $X_{t,i}$ at time $t$ and $X_{t+1,j}$ at time $t+1$. The matrix $\hat{\boldsymbol{\pi}}$ can therefore be interpreted as an estimated coupling between the empirical distributions at the two time points. Each entry $\hat{\pi}_{ij}$ represents the amount of probability mass transported from cell $i$ at time $t$ to cell $j$ at time $t+1$, under the assumption that cells evolve in the state space in a least-effort manner. In this sense, $\hat{\boldsymbol{\pi}}$ provides a probabilistic description of the developmental transitions between cellular states across consecutive time points. 
However, there are two limitations with this vanilla OT approach. First, numerically solving the OT on empirical distributions is computationally expensive. An entropic penalization is often introduced to the original OT objective function Eq. \eqref{vanillaot} as follows:
\begin{equation}\label{EOT}
\begin{aligned}
{\boldsymbol{\hat\pi}}
= &\arg\min_{\boldsymbol{\pi} \in \mathbb{R}_+^{N_t\times N_{t+1}}}
\Bigg\{
 \sum_{i=1}^{N_t}\sum_{j=1}^{N_{t+1}} \pi_{ij}\, \|X_{t,i}-X_{t+1,j}\|_2^2 
\Bigg\}+ \lambda \sum_{ij}\pi_{ij}(\log\pi_{ij}-1),
\\ &\text{s.t.} \sum_{i=1}^{N_{t}} \pi_{ij} = \frac{1}{N_{t+1}}, ~~\sum_{j=1}^{N_{t+1}} \pi_{ij} = \frac{1}{N_{t}},~~\pi_{ij}\geq 0, 
% \\
% & \text{where} \qquad r_i := \sum_{j=1}^{N_{t+1}}\Pi_{ij} \qquad\text{and}\qquad
% c_j := \sum_{i=1}^{N_t}\Pi_{ij}. \qquad
\end{aligned}
\end{equation}
where $   \sum_{ij}\pi_{ij}(\log\pi_{ij}-1)$ is the entropy of $\boldsymbol{\pi}$ and $\lambda>0$ is a hyperparameter. 
The KKT conditions given by the entropic OT  induce a fast iterative matrix-scaling approach on real data called Sinkhorn algorithm \citep{sinkhorn1967diagonal,cuturi2013sinkhorn}. Larger $\lambda $ encourages that $\pi_{{P}, {Q},\lambda}$ moving closer to the uniform joint distribution, i.e., the product measure of  the two empirical  marginal  distributions and $\lim_{\lambda\to+\infty} \hat\pi_{ij} =\frac{1}{N_t N_{t+1}}$. When $\lambda\to0^+$, the solution given by Eq. \eqref{EOT} will converge to the vanilla OT solution defined by \eqref{vanillaot},  but may encounter numerical issues or large computational cost.  
Secondly, cells will naturally undergo proliferation and apoptosis (cellular growth and death), which are not considered by the vanilla OT framework, especially the growth rate of cells may vary in different cells and unknown. One approach adopted by prior works  to incorporate the growth rate  is to assume that cells follow a two-phase procedure that cells first undergo the growth then differentiate or transit cell states following the optimal transport principle \citep{schiebinger2019optimal,zhang2021optimal,klein2025mapping}, as conceptualized in the panel (a) of Figure \ref{Fig.main}.  Let $g(x) \ge 0$ denote the relative growth rate of cells at biological state $x$, and let $Q_t$ denote the marginal distribution after cellular growth, defined such that
$\frac{d {Q}_t}{d {P}_t}(x) =\frac{g(x)}{\int g(x)d {P}_t(x)}$, where $\frac{d {Q}_t}{d {P}_t}(x)$ is the Radon–Nikodym derivative.  The relative growth rate $g(x)$ is typically unknown. Previous work has attempted to estimate it using models such as birth–death processes combined with curated gene sets associated with proliferation and apoptosis, although different studies adopt different strategies to give a prior  estimate  of  $g(x)$. Given an estimated growth rate $\tilde g(x)$,  existing approaches such as Waddington-OT and Moscot  then consider the following unbalanced optimal transport problem:
% \begin{equation}\label{UOT}
% \begin{aligned}
% \hat \pi
% =\arg\min_{\pi\in \mathcal{M}_+(\mathbb{R}^D\times\mathbb{R}^D)}
% \Bigg\{
% &\int_{\mathbb{R}^D \times \mathbb{R}^D}\|x-y\|_2^2\, d\pi(x,y)
% - \lambda_0\,\text{E}(\pi) 
% \\
% &\quad + \lambda_1\,H\!\left(\pi_x \,\vert \widehat{{Q}}_{t}\right)
% + \lambda_2\,H\!\left(\pi_y \,\vert \widehat{{P}}_{t+1}\right)
% \Bigg\},
% \end{aligned}
% \end{equation}
% where $ \mathcal{M}_+(\mathbb{R}^D\times\mathbb{R}^D) $ is the space of probability distributions on $\mathbb{R}^D\times\mathbb{R}^D$, $\pi_x$ and $\pi_y$ are the marginal distributions of $\pi$ and $H\!\left(\cdot \,\middle\|\,\cdot \right)$ is the KL divergence between two distributions.
\begin{equation}\label{UOT_discrete_sum}
\begin{aligned}
\hat{\boldsymbol{\pi}}
= &\arg\min_{\boldsymbol{\pi} \in \mathbb{R}_+^{N_t\times N_{t+1}}}
\Bigg\{
 \sum_{i=1}^{N_t}\sum_{j=1}^{N_{t+1}} \pi_{ij}\, \|X_{t,i}-X_{t+1,j}\|_2^2 
+ \lambda_0 \sum_{i=1}^{N_t}\sum_{j=1}^{N_{t+1}}\pi_{ij}
\left(\log {\pi_{ij}}-1\right)
\\
&\quad + \lambda_1 \sum_{i=1}^{N_t} r_i \log\!\left(\frac{r_i}{a_i}\right)
+ \lambda_2 \sum_{j=1}^{N_{t+1}} c_j \log\!\left(\frac{c_j}{b_j}\right)
\Bigg\},
% \\
% & \text{where} \qquad r_i := \sum_{j=1}^{N_{t+1}}\Pi_{ij} \qquad\text{and}\qquad
% c_j := \sum_{i=1}^{N_t}\Pi_{ij}. \qquad
\end{aligned}
\end{equation}
\text{where} $\pi_{ij}\geq 0$ is the $(i,j)$-entry of the non-negative matrix $\boldsymbol{\pi}$, $a_i = \tilde{g}(X_{t,i})/\sum_{k=1}^{N_t}\tilde{g}(X_{t,k})$ is the growth-rate adjusted marginal distribution at time $t$, $b_j = 1/N_{t+1} $ is the marginal distribution at time $t+1$,  and  $ r_i := \sum_{j=1}^{N_{t+1}}\pi_{ij} $ and $
c_j := \sum_{i=1}^{N_t}\pi_{ij}$ are the row sum and the column sum of $\pi$, respectively. 
 The use of KL penalties such as $\lambda_1 \sum_{i=1}^{N_t} r_i \log\!\left(\frac{r_i}{a_i}\right)
+ \lambda_2 \sum_{j=1}^{N_{t+1}} c_j \log\!\left(\frac{c_j}{b_j}\right)$  in Eq. \eqref{UOT_discrete_sum}  in place of hard marginal constraints in   Eq. \eqref{EOT}  is motivated by the fact that they can potentially account for misspecification of estimated growth model $\tilde{g}(x)$ . The soft KL penalties instead allow controlled deviations from the empirical marginals, with the tuning parameters $\lambda_1$ and $\lambda_2$ governing the strength of this relaxation.  

However, unbalanced OT with prior-estimated growth rates still faces several important challenges. First, even when prior knowledge from curated sets of growth-related genes is incorporated, the inferred growth patterns often appear overly sparse or exhibit inflated variance \citep{halmos2025dest}. In particular, even within the same cell type, only a small number of cells may be assigned large growth rates, whereas most others are inferred to have nearly vanishing growth or even cell death. This behavior may be caused by  misspecification of the growth-rate model or  computational artifact. Second, the optimal transport framework is formulated for the uncontaminated distribution $P_t$, whereas in practice observed cellular features are subject to technical noise \citep{sarkar2021separating}. Moreover, despite the inclusion of an entropic penalty, solving the unbalanced OT problem in Eq.~\eqref{UOT_discrete_sum} remains computationally demanding and may fail to converge in a reasonable amount of time unless low-rank approximations or more advanced computational resources are employed \citep{klein2025mapping}.

To address these concerns, we propose a TIMO that combines the metacell concept with semi-relaxed optimal transport. Before introducing our method, we first review the background on metacells.

\subsection{Metacell}
Metacell were first proposed by \cite{baran2019metacell},  representing homogeneous  groups of cell that the within group variation is considered to be technical noise or sampling effect in the sequencing platform. 
The main advantage of metacell is  that it can   enhance biological signal  and reducing the computational burden in large-scale scRNA-seq data. Following the data-generating schemes considered in \cite{sarkar2021separating} and \cite{liu2025mcrigor}, we model individual cells using a mixture model framework, in which metacells correspond naturally to mixture components. 
For notational simplicity, we consider a general setting that  we have $n$ cells and $D$ features, where $D$ may denote either the number of measured genes or the dimension of a low-dimensional representation. Suppose there are $m$ metacells (or mixture components), and that the $n$ cells are independently generated according to $   S_i \sim \mathrm{Multi}(p_1,\ldots,p_m)$ and  $X_i \mid S_i=k \sim F_{k},$
 where $F_{k}$ is a probability distribution with  bounded support.
% \kong{Notation $F_i$ and $F_{S_i}$ needs to be consistent. In addition, is diagonal covariance matrix assumption too restrictive here?} \zhu{ The assumption of a diagonal covariance matrix is made for technical convenience in the  proof  and can be relaxed without significant additional difficulty. I can remove this assumption. } \kong{Yes, then it would be better to just list the assumptions in the theoretical part.}
Although the metacell concept is closely related to clustering and mixture modeling, it is distinct from the notion of cell type. Cell types are typically defined using prior biological knowledge, whereas a metacell represents a much smaller and more homogeneous groups of cells.  As a result, multiple metacells may be nested within the same cell type, allowing statistical analysis at metacell resolution to capture, rather than overlook, heterogeneity within cell types.    For example, the Metacell \citep{baran2019metacell} and Metacell-2 \citep{ben2022metacell} algorithms construct a k-nearest-neighbor (kNN) graph based on gene expression profiles and then partition the graph using divide-and-conquer strategies. 
More recently, SEACells \citep{persad2023seacells} was proposed to identify metacells using a kNN graph built on low-dimensional cell embeddings together with nonnegative matrix factorization. This approach is also compatible with scATAC-seq and single-cell multi-omics data. In our application, we use SEACells \citep{persad2023seacells} to map cells to metacells because it operates in latent embedding space and is particularly well suited to our setting. Specifically, we study the large atlas  data from \citet{qiu2022systematic} which combines scRNA-seq data from multiple sources, so batch correction based on low-dimensional representations, such as Seurat \citep{butler2018integrating} or Harmony \citep{korsunsky2019fast}, is required. A  visualization example  of metacell assignment on E7.0 cells from \citet{qiu2022systematic} is provided in Figure \ref{E7}. 
For a broader and more detailed overview of metacell methods, we refer the reader to the review article by  \cite{bilous2024building}.

\subsection{TIMO}\label{sec:timo}
With the necessary background in place, we now introduce our proposed method TIMO. For simplicity, suppose that at time $ t$ we observe cells    $\{X_{t,i}\}_{i=1}^{N_t}\subset\mathbb{R}^D$,   where $D$ is the dimension of low-dimensional embedding space obtained via Seurat \citep{butler2018integrating}. In our case study, we set  $D=30$, which is a common choice in scRNA-seq data analysis.   We first apply SEACells algorithm to  $\{X_{t,i}\}_{i=1}^{N_t}\subset\mathbb{R}^D$ for each time point separately. We assign each cell a metacell label $S_{t,i}\in\{1,\ldots,m_t\}$, where $m_t$ denotes the number of metacells at time $t$.
Since cells within the same metacell are constructed to be homogeneous in their biological state, we regard within-metacell variability as technical noise introduced by the sequencing platform rather than genuine biological variation. Consequently, cells belonging to the same metacell are assumed to represent a common underlying biological state and, therefore, to share the same  distribution at the previous time point as well as the same transition probabilities to biological states at the subsequent time point.  This motivates imposing an invariance constraint on the transport plan: for any source metacell $k$, all cells $i$ with $S_{t,i}=k $ share identical transport probabilities to each target metacell $\ell$; equivalently, the coupling is constant within each   metacell block.

Formally, let $ S_{t,i}\in\{1,\dots,m_t\}$ and $S_{t+1,j}\in\{1,\dots,m_{t+1}\}$ denote metacell labels. Consider the single-cell level coupling $\pi\in\mathbb{R}_+^{N_t\times N_{t+1}}$ under the block-constant constraint
$$\pi_{ij} \;=\; \gamma_{S_{t,i},\,S_{t+1,j}}
\quad\text{for some } \gamma\in\mathbb{R}_+^{m_t\times m_{t+1}}.$$
Under this constraint, the single-cell transport objective defined by squared Euclidean cost decomposes into between-metacell and within-metacell components:
$$\sum_{i=1}^{N_t}\sum_{j=1}^{N_{t+1}} \pi_{ij}\,\|X_{t,i}-X_{t+1,j}\|_2^2\,
\propto\,
\sum_{k=1}^{m_t}\sum_{\ell=1}^{m_{t+1}}
 \gamma_{k\ell}\,
\Big\|\hat\mu_{t,k}-\hat\mu_{t+1,\ell}\Big\|_2^2
\;+\; \text{const},$$
where $\hat\mu_{t,k}$ and $\hat\mu_{t+1,\ell}$ are metacell centroids, and the additive term  only aggregates within-metacell variation. This identity is a weighted version of the standard ANOVA decomposition of squared distances into between-group and within-group variation. As a result, optimizing OT over block-constant couplings at the single-cell level is equivalent to optimizing a metacell-level OT problem with cost defined by squared distances between metacell centroids. Therefore, with the metacell label, 
       we   calculate the centroids of each metacell, and obtain $$\hat{p}_{t,k}= \frac{\sum_{i=1}^{N_t}  I\{S_{t,i}=k\}}{N_t},~~\text{and}~~\hat\mu_{t,k} =\frac{1}{N_t\hat{p}_{t,k}} \sum_{i=1}^{N_t}X_{t,i}\cdot  I\{S_{t,i}=k\}. $$
For any two consecutive time points $t$ and $t+1$, we solve the following semi-relaxed optimal transport problem: 
% \begin{equation}\label{semi_relaxed_ot}
% \begin{aligned}
% \hat{\omega}t
% =\arg\min{\omega\in\mathbb{R}+^{m_t\times m{t+1}}}
% \Bigg{
% &\sum_{k=1}^{m_t}\sum_{\ell=1}^{m_{t+1}}
% \omega_{k\ell}, |\hat\mu_{t,k}-\hat\mu_{t+1,\ell}|_2^2
% 	•	\varepsilon \sum_{k=1}^{m_t}\sum_{\ell=1}^{m_{t+1}}
% \omega_{k\ell}\log!\left(\frac{\omega_{k\ell}}{\hat p_{t,k}\hat p_{t+1,\ell}}\right)
% \
% &\quad + \lambda_1 \sum_{k=1}^{m_t} r_k \log!\left(\frac{r_k}{\hat p_{t,k}}\right)
% 	•	\lambda_2 \sum_{\ell=1}^{m_{t+1}} c_\ell \log!\left(\frac{c_\ell}{\hat p_{t+1,\ell}}\right)
% \Bigg},
% \end{aligned}
% \end{equation}
\begin{equation}\label{SOT}
\begin{aligned}
\boldsymbol{\widehat{\pi}^{(t,t+1)}}
= &\arg\min_{\boldsymbol{\pi} \in \mathbb{R}_+^{m_t\times m_{t+1}}}
\Bigg\{
 \sum_{k=1}^{m_t}\sum_{\ell=1}^{m_{t+1}} \pi_{k\ell}\, \|\hat\mu_{t,k}-\hat\mu_{t+1,\ell}\|_2^2 
+ \lambda_0 \sum_{k=1}^{m_t}\sum_{\ell=1}^{m_{t+1}}\pi_{k\ell}
\left(\log {\pi_{k\ell}}-1\right)
\\
&\quad + \lambda_1 \sum_{k=1}^{m_t} \left(\sum_{\ell=1}^{m_{t+1}} \pi_{k\ell}\right)\log\!\left(\frac{\sum_{\ell=1}^{m_{t+1}}\pi_{k\ell}}{\hat{p}_{t,k} }\right)
\Bigg\},
\\&~\text{s.t.} \sum_{k=1}^{m_{t}} \pi_{k\ell} = \hat{p}_{t+1,\ell},~~ \pi_{k,\ell}\geq 0. 
\end{aligned}
\end{equation}
Then the estimated growth rate for each metacell at time $t$ is given by 
\begin{equation}\label{gr}
    \widehat{g}_{t,k} = \sum_{\ell=1}^{m_{t+1}} \widehat\pi^{(t,t+1)}_{k,\ell}/\widehat{p}_{t,k}. 
\end{equation} 
The optimization problem in Eq.~\eqref{SOT} can be efficiently solved using a generalized Sinkhorn-type algorithm, as described in Section S1 of the Supplementary Materials.  The convergence to the optimal solution under some mild conditions  has been theoretically established for this class of algorithms. We refer readers to \cite{robot} and \cite{fukunaga2022convergence} for comprehensive study  of the corresponding algorithmic convergence.
The conceptual visualization of TIMO is also provided in Figure \ref{Fig.main}. 
TIMO differs from Waddington-OT and Moscot in two main respects. First, TIMO does not rely on a pre-specified growth-rate model constructed from curated gene sets. Second, TIMO employs a semi-relaxed optimal transport formulation, rather than the fully unbalanced formulation adopted in Waddington-OT and Moscot.
These design choices are motivated by both statistical and practical considerations. In existing approaches, growth-rate estimates are typically derived from curated gene sets or other sources of external biological knowledge. However, such gene sets may be incomplete, platform-dependent, or only partially observed in sequencing data. When the growth-rate model is misspecified, enforcing it through strong marginal relaxation can bias the transport plan toward incorrect mass redistribution. In particular, an inaccurate growth prior may drive the coupling away from biologically plausible transitions.
In contrast, TIMO avoids imposing an explicit growth model and aims to learn the unknown growth rates in an ``unsupervised'' manner, allowing deviations that can accommodate various   growth rates between consecutive time points. This semi-relaxed formulation captures potential distributional changes attributable to proliferation or death without forcing them to follow a predefined model. We retain the marginal constraint at time $t+1$ as a hard constraint. The asymmetric treatment of the marginal constraints relaxing the constraint at time $t$ while enforcing a hard constraint at time $t+1$   is motivated by both biological and statistical considerations.   From a biological perspective, cells observed at time $t+1$ must originate from some population of cells at time $t$, and therefore each individual cell observed at time $t+1$ must be explained by ancestors at the earlier time point $t$ or equivalently the distribution at time $t+1$ should be fully preserved in the transport plan. In contrast, cells present at time $t$ do not necessarily persist to time $t+1$, as some cells may die.  From a statistical perspective, after aggregating cells into metacells, we regard the empirical metacell proportions at time $t+1$ as a stabilized estimate of the distribution over biological states represented by the metacell centroids. Because metacell construction serves as a denoising procedure, we assume that stochastic fluctuations in the centroid representation are negligible relative to the biological signal. Under this assumption, enforcing the target marginal constraint ensures that the inferred transport plan aligns with the observed distribution of cellular states  at time $t+1$. 

The semi-relaxed objective also implicitly admits an equivalent marginal representation that further clarifies its statistical interpretation. Let $\Delta^{K-1} = \{\boldsymbol{p} \in \mathbb{R}^K : \sum_{k=1}^K p_k = 1,\; p_k > 0\}$ denote the $(K-1)$-dimensional probability simplex. Let $\boldsymbol{r} = (r_1, \ldots, r_{m_t}) \in \Delta^{m_t-1}$ be a probability vector. Let $\boldsymbol{\hat{p}}_t = (\hat{p}_{t,1}, \ldots, \hat{p}_{t,m_t}) \in \Delta^{m_t-1}$ and $\boldsymbol{\hat{p}}_{t+1} = (\hat{p}_{t+1,1}, \ldots, \hat{p}_{t+1,m_{t+1}}) \in \Delta^{m_{t+1}-1}$ denote the empirical distributions of metacell proportions at time points t and t+1, respectively.
Let $\mathrm{W}_{\lambda_0}(\boldsymbol{r},\boldsymbol{\hat{p}_{t+1}})$ be the entropically regularized version of squared Wasserstein distance between marginal distributions representing by vector $\boldsymbol{r}$ and $\boldsymbol{\hat{p}_{t+1}}$ as follows: 
% \kong{Need to define $\boldsymbol{r}$ and $\boldsymbol{\hat{p}_{t+1}}$, especially the objective function does not contain $\boldsymbol{r}$ and $\boldsymbol{\hat{p}_{t+1}}$ for the following equation}
\begin{equation}\label{Edist}
\mathrm{W}^2_{2,\lambda_0}(\boldsymbol{r},\boldsymbol{\hat{p}_{t+1}})
:=
\min_{\pi\in \Pi(\boldsymbol{r},\boldsymbol{\hat{p}_{t+1}})}
\left\{
\sum_{k=1}^{m_t}\sum_{\ell=1}^{m_{t+1}} \pi_{k\ell}\,
\|\hat\mu_{t,k}-\hat\mu_{t+1,\ell}\|_2^2
+
\lambda_0 \sum_{k=1}^{m_t}\sum_{\ell=1}^{m_{t+1}}
\pi_{k\ell}(\log \pi_{k\ell}-1)
\right\}.
\end{equation}
By partial minimization with respect to the coupling, problem \eqref{SOT} can be written as
\begin{equation}\label{marginal}
    \min_{\boldsymbol{r} \in \Delta^{m_t-1}}
\left\{
\mathrm{W}^2_{2,\lambda_0}(\boldsymbol{r}, \boldsymbol{\hat p_{t+1}})
+
\lambda_1 \mathrm{KL}(\boldsymbol{r}\| \boldsymbol{\hat p_t})
\right\},
\end{equation}
where $\boldsymbol{r}$ represents the post-growth marginal distribution of metacells at time $t$, and $\Delta^{m_t-1}$ denotes the probability simplex.
In single-cell developmental data, without longitudinal tracking, growth effects cannot be uniquely disentangled from state transitions.  From this perspective, the KL term defines an information-theoretic uncertainty set around the observed distribution $\boldsymbol{\hat p_t}$.  Equivalently, the penalized problem corresponds to minimizing  transport cost over a KL-divergence ball centered at $\boldsymbol{\hat p_t}$. Thus, TIMO can be also interpreted as finding a  marginal distribution of metacell at time $t$ that is allowed to vary within an uncertainty region induced by unknown  growth-rate, while simultaneously minimizing the  transport cost to the observed distribution at the subsequent time point. 

\section{Real Data Analysis}
\label{sec:data}
In this section, we  apply TIMO  to the mouse embryogenesis atlas data containing almost 1.7 million cells integrated by \cite{qiu2022systematic}.    This dataset contain single-cell snapshot measurements collected across 20 successive developmental stages, spanning embryonic day E3.5 to E13.5. The aim of this large-scale single-cell dataset is to study the cell-state trajectories of early mouse embryo development.  Before applying TIMO, we preprocess the data following the pipeline in \cite{klein2025mapping}.  

\begin{figure}[]
\centering  
\includegraphics[width=0.95\textwidth]{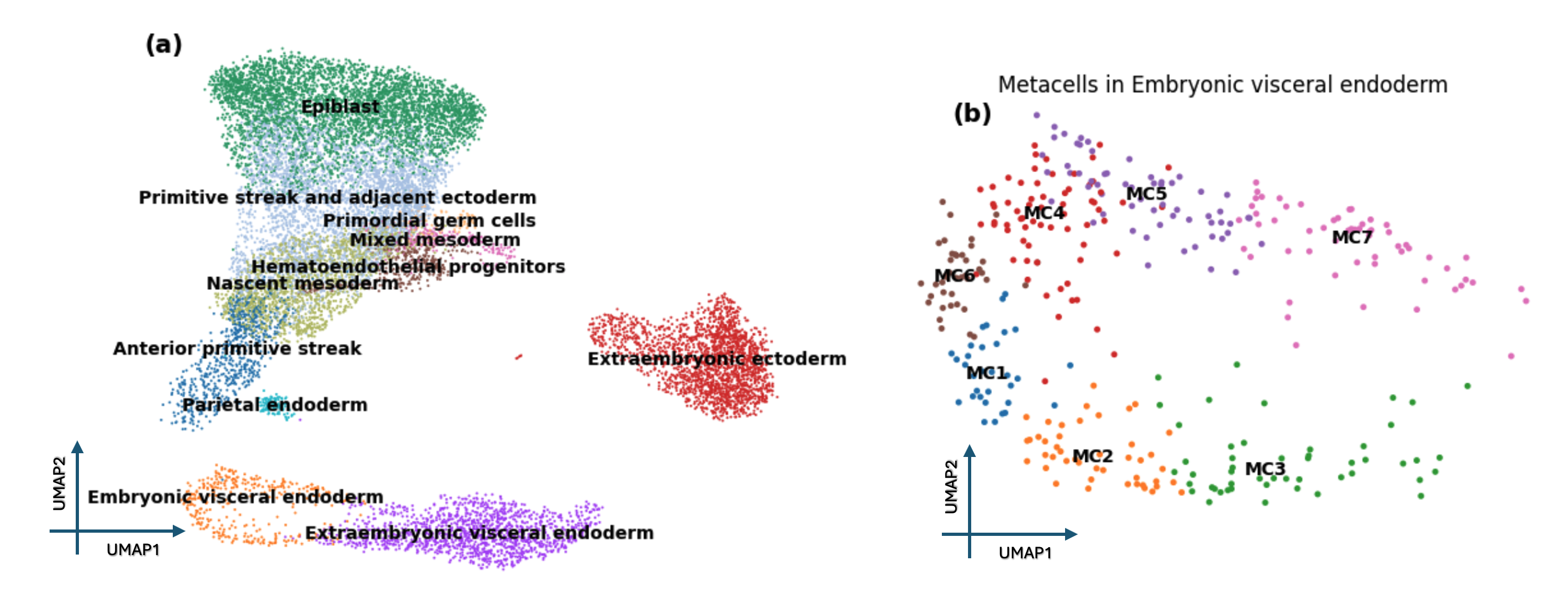}  
\caption{\textbf{Example of metacell construction at E7.0.}
\textbf{(a)} UMAP projection of E7.0 cells, where colors indicate priorly annotated cell types.
\textbf{(b)} Metacell assignments within the embryonic visceral endoderm. Colors denote different metacells constructed based on cellular feature similarity through SEACells \citep{persad2023seacells}.}
\label{E7}
\end{figure}

We begin by grouping cells into metacells using SEACells \citep{persad2023seacells}. An important hyperparameter in metacell construction is the desired level of granularity, defined as the ratio of the number of individual cells to the number of metacells. For example, a graining level of 10 means that, on average, each metacell contains 10 cells. A recent review and protocol on metacell methods recommends graining levels between 10 and 75 \citep{bilous2024building}. Following this guidance, we choose the graining level adaptively according to the sample size $N_t$. Specifically, we set it to 10 when $N_t < 100$, 20 when $100 \leq N_t < 1{,}000$, 30 when $1{,}000 \leq N_t < 10{,}000$, 50 when $10{,}000 \leq N_t < 100{,}000$, and 75 when $N_t \geq 100{,}000$. In addition, we also follow the suggestion in \cite{bilous2024building} to build metacells in a ``supervised'' manner when cell-type labels are available by constructing metacells for each cell type separately. This strategy can tackle the issue that rare-cell types is possible to be  aggregated with other major cell types   and be   completely missed for down-stream analysis.  Note that this similar strategy of constructing metacell is also adopted by another advanced  analysis of the evolution of pancreatic tumors \citep{burdziak2023epigenetic}. In Figure \ref{E7}, we visualize the E7.0 cells and their corresponding metacell assignments in the embryonic visceral endoderm from the atlas dataset to illustrate the concept of metacells. As shown in the figure, even within the same cell type, cells are partitioned into multiple metacells based on their feature similarity, reflecting finer-scale heterogeneity in the data.  

Using the metacell assignments obtained from SEACells, we solve the semi-relaxed optimal transport problem in Eq.\eqref{SOT} for each consecutive time pair $(t,t+1)$, and compute the associated metacell-level growth rates according to Eq.\eqref{gr}.
 Since the objective function in Moscot is identical to Waddington-OT in this temporal problem and is  a more computationally efficient  implementation with reproducible instructions, we compare the results between TIMO and Moscot for convenience. In the comparison, we set $\lambda_0 =0.005$, and $\lambda_1 =0.1$ in Eq.\eqref{SOT} throughout all time points.
To obtain the Moscot results, we carefully follow the authors’ implementation and reproduce their results using the code available in their reproducibility repository. Choosing hyperparameters remains a challenging problem for optimal transport methods, Moscot and Waddington-OT generally set  the entropic parameter to be numbers between $0.005$ and $0.05$ and the unbalance parameter between $0.1$ and $1.0$ and select them based on several benchmark metrics. While the TIMO hyperparameters could in principle also be selected in this way, we instead hold them fixed across all time points to ensure a fair comparison.

\subsection{Cell-type Transition Estimate}\label{sec:transition}
For any pair of cell types A and B, we compute the transition probability by aggregating the entries of $\boldsymbol{\hat{\pi}_t}$ whose rows correspond to metacells belonging to cell type A and whose columns correspond to cell type B. Visualization of the  transition probabilities are presented in Section S5.  While most of the cell-type transitions identified by TIMO agree with the developmental atlas in \cite{qiu2022systematic}, we find several additional differentiation events that were not reported in the original atlas. For example, we find that epiblast cells from E6.25 to E7.0 consistently give rise to primitive streak and adjacent ectoderm at the subsequent time point, whereas \cite{qiu2022systematic} reports this transition at E6.25 and E6.75 but not at the intermediate stage E6.5. This result suggests a more continuous differentiation process than that implied by the original atlas.  In addition, we find that gut cells at E7.5-E7.75 may have potential ancestors in visceral endoderm from earlier stages, consistent with the mosaic nature of gut endoderm and with the open question noted in \cite{nowotschin2020guts} as to the extent of visceral endoderm contribution to later gut tube derivatives.

We then benchmark TIMO against Moscot using the estimated cell-type transition probabilities to show TIMO can provide statistical benefits. As Moscot also provides a low-rank approximation implementation to improve computational efficiency, we additionally include this variant in the comparison. 
Similar aggregations are also applied to  the single-cell level transport matrix given by Moscot and the low-rank variant of Moscot. 
Two benchmark metrics proposed in \cite{klein2025mapping} are used to evaluate the accuracy of the inferred cell-type transitions: a germ-layer based metric and a curated cell-type metric.
The germ-layer metric is motivated by the biological prior knowledge that cells rarely transition across germ layers during development. Transitions occurring within the same germ layer are regarded as correct, while transitions across different germ layers are considered incorrect. The curated cell-type metric evaluates the estimated transition probabilities in a curated reference set collected from previously reported cell-type transitions in massive existing literature. Under this metric, transitions matching the curated set are treated as correct, while those that do not match are considered incorrect.

\begin{figure}[H]
\centering  
\includegraphics[width=0.75\textwidth]{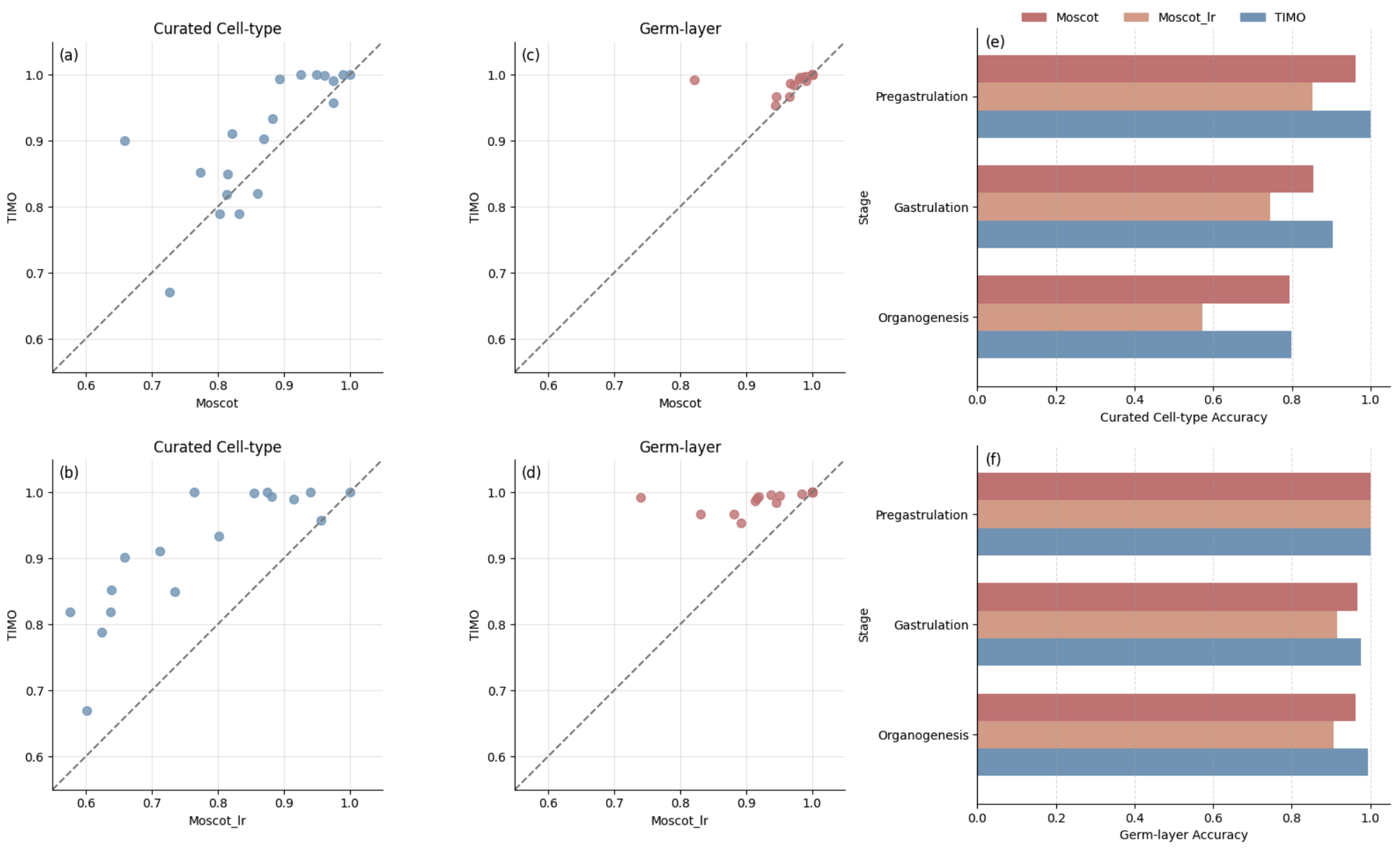}  
\caption{\textbf{Benchmark of cell-type transition estimation accuracy.}
\textbf{Panels (a)–(d):} Scatter plots comparing the accuracy of cell-type transition estimates produced by TIMO, Moscot, and Moscot with low-rank approximation across all time points in the mouse embryo developmental atlas dataset. TIMO consistently outperforms both Moscot variants in terms of accuracy on curated cell-type transitions and germ-layer transitions across the 19 time points.
\textbf{Panels (e)–(f):} Weighted accuracy of TIMO, Moscot, and Moscot with low-rank approximation across three developmental stages. }
\label{ME_atlas_scatter}
\end{figure}

We evaluate these two types of accuracy metrics on TIMO, Moscot and low-rank Moscot  and visualize the comparisons by scatter plots in panels (a)-(d) of  Figure \ref{ME_atlas_scatter}. Across all consecutive time-point pairs, TIMO consistently outperforms both variants of Moscot in estimating cell-type transitions in terms of  curated cell-type accuracies and germ-layer based accuracies. 
Moreover, we also compare the accuracies based on three different developmental stages, where time points before E6.5 belong to the pregastrulation stage, time points between E6.5 and E8.5 belong to the gastrulation stage and time points later than E8.5 are considered as organogenesis stages.  Within each stage, we compute a weighted average of the accuracies, with weights proportional to the number of cell types present at each time point. TIMO consistently achieves higher accuracy  across all three stages. Particularly, during the gastrulation stage, TIMO achieves a substantially higher  accuracy (0.904) compared to Moscot (0.854) on curated cell types. This stage corresponds to a critical developmental period when pluripotent epiblast cells diversify into the three germ layers (ectoderm, mesoderm, and endoderm) and multiple early progenitor populations.

We also emphasize  that bootstrap  can be used to perform valid uncertainty quantification for the cell-type transition probabilities estimated by TIMO, with theoretical support for bootstrap consistency provided in Theorem \ref{cor1} in Section \ref{sec:theory}. This approach is computationally feasible because TIMO solves OT on metacell-level summary statistics instead of the original single-cell level data. We use $500$ bootstrap resamples to construct $95\%$ confidence intervals for the 10 largest observed cell-type transition entries between the E7.0 and E7.25 stages. The results are presented in Section S5 and show that in real data the limiting distribution of the TIMO estimator is well approximated by both the bootstrap and normal distributions, despite the theoretical justification relying on several assumptions.

\vspace{-0.75cm}

\subsection{Comparison on Growth Rate Estimate}\label{sec:growth}
\begin{figure}[H]
\centering  
\includegraphics[width=1.01\textwidth]{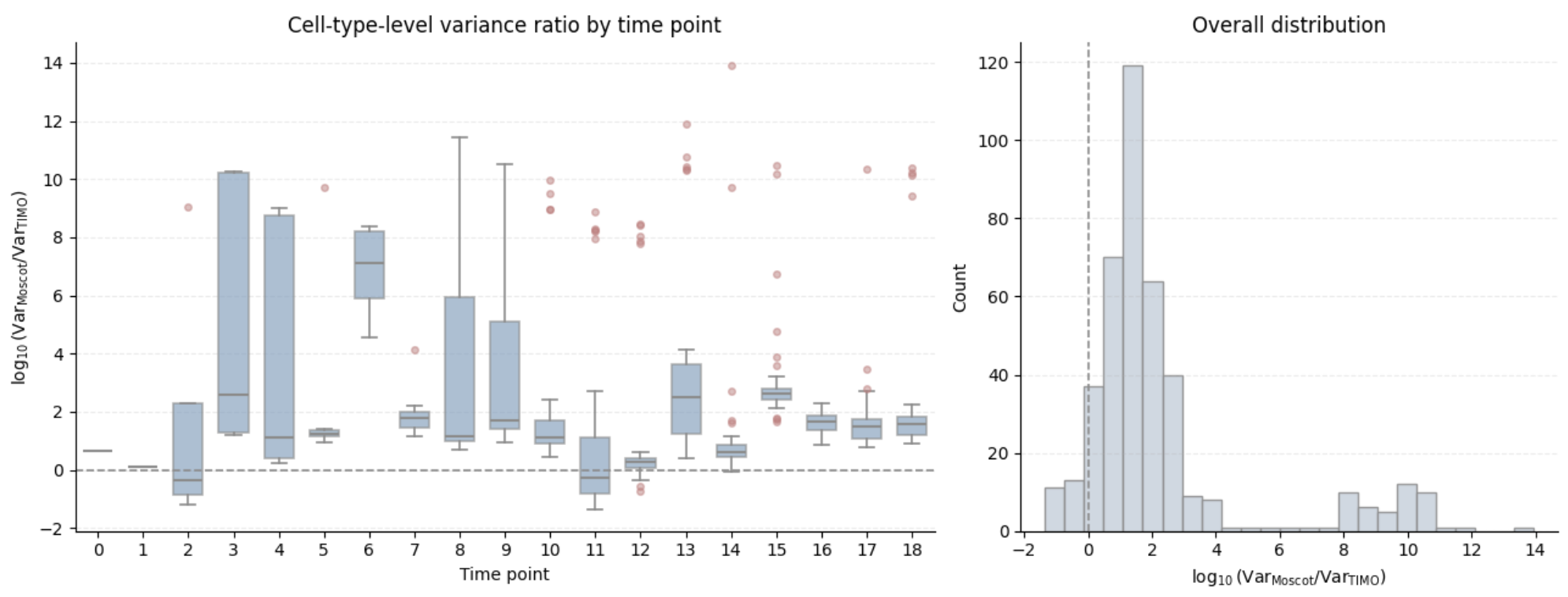}  
\caption{\textbf{Comparison of cell-type-level growth-rate variance between Moscot and TIMO across developmental time points.} For each time point and each annotated cell type, we computed the variance of cell-level growth-rate estimates under TIMO and Moscot, and summarized their relative variability by the log-variance ratio,$ \log_{10}(\mathrm{Var}_{\mathrm{Moscot}}/\mathrm{Var}_{\mathrm{TIMO}}$).
\textbf{Left:} boxplots of the log-variance ratio across cell types at each time point. \textbf{Right:} overall distribution pooled across all cell types and time points. Positive values indicate larger within-cell-type variance under Moscot, whereas negative values indicate larger variance under TIMO. Across most time points, the log-variance ratio is predominantly positive, indicating that TIMO generally yields lower within-cell-type growth-rate variance and therefore more stable growth-rate estimates than Moscot. The dashed horizontal/vertical line marks 0, corresponding to equal variance between the two methods.}
\label{gr_var}
\end{figure}

We then compare the estimated growth rates obtained from TIMO and Moscot.  Note that Moscot estimates their posterior growth rates using a similar formulation to Eq.~\ref{gr}, but applies it directly to the single-cell transport matrix.
 To evaluate the biological validity  of the estimated growth rates, we adopt the so-called growth distortion metric proposed in a recent study of spatial-temporal transcriptomics \citep{halmos2025dest}, which is mathematically equivalent to  the  within-cell-type variances of the estimated growth rates. A smaller distortion indicates a more reliable estimate, since cells within the same cell type are expected to exhibit relatively homogeneous or smoothly varying growth rates, rather than highly concentrated values in a small subset of cells while remaining close to zero for most cells in the same cell type \citep{milan1996cell}. For each time point and each cell type, we first compute the within-cell-type variance of cell-level growth-rate estimates and compare the two methods through the variance ratio $\mathrm{Var}_{\mathrm{Moscot}} / \mathrm{Var}_{\mathrm{TIMO}}$, as shown in Figure \ref{gr_var}. Across most developmental time points, TIMO yields lower within-cell-type growth-rate variance than Moscot.

\begin{figure}[H]
\centering  
\includegraphics[width=0.9\textwidth]{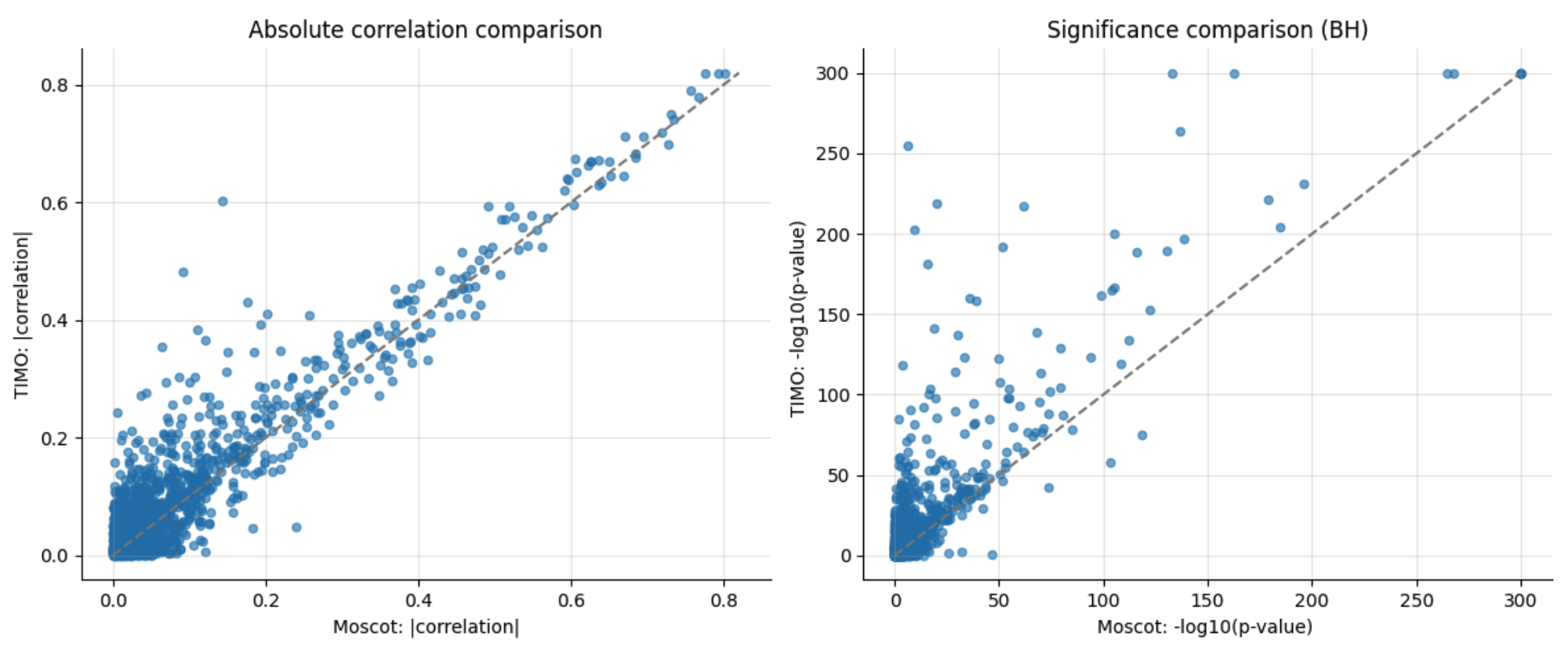}  
\caption{\textbf{Comparison of growth-rate associations with curated gene sets between TIMO and Moscot.}
\textbf{Left:} scatter plot comparing the absolute Spearman correlations between estimated growth rates and curated gene sets obtained by TIMO (y-axis) and Moscot (x-axis). 
\textbf{Right:}  scatter plot comparing the corresponding statistical significance measured by $-\log_{10}(p\text{-value})$ after BH correction. 
The dashed line represents the equality line $y=x$. Points lying above the diagonal indicate stronger correlations or higher statistical significance for TIMO. 
Across most gene sets, TIMO exhibits larger correlations and higher significance levels than Moscot, suggesting that TIMO provides more biologically meaningful estimates of cellular growth rates.}
\label{cor_scatter}
\end{figure}

We also conduct an association based analysis on the growth rate. Since there is no ground truth for the relative growth rate, one way to assess the effectiveness of the estimated growth rates is to test the association between the estimated growth rates and the expression levels of growth-related genes.  A better estimate of the growth rate should have stronger correlations with the  genes associated with cell growth. We set the growth rate of  each individual cell to be the growth rate of its corresponding metacell estimated by TIMO. Then we calculate the Spearman correlation between a pre-defined set of growth-related genes, where this gene set contains 281 genes and  was used in Moscot to give a prior estimate of the growth rates. Overall, while the two methods identify a broadly similar number of significant markers across most developmental time points, TIMO recovers substantially more markers during the E6.5–E7.5 interval, corresponding to the core gastrulation stage. Specifically, TIMO identifies 183, 156, 202, 208, and 170 genes at E6.5, E6.75, E7.0, E7.25, and E7.5, respectively, whereas Moscot identifies 161, 88, 151, 203, and 146 genes at the same time points under the Benjamini–Hochberg correction with a false discovery rate (FDR) threshold of 0.05. For instance, TIMO identifies \textit{Rb1}, a key regulator of cell fate 
\noindent that coordinates cell-cycle progression and apoptosis \citep{indovina2015rb1}, whereas Moscot does not detect this gene at any of these five time points.
We also visualize the Spearman correlations between the estimated growth rates and curated proliferation- and apoptosis-related gene sets in Figure \ref{cor_scatter}. The results show that during this developmental interval, the growth rates estimated by TIMO exhibit stronger correlations with these gene sets than those estimated by Moscot, even though TIMO does not incorporate such prior biological information. This developmental interval is highly dynamic, characterized by rapid proliferation, extensive cell fate transitions, and major tissue reorganization. Previous studies have also reported increased apoptosis and cell clearance during gastrulation \citep{abraham2024single}. Because these markers are identified based on the growth rates estimated by TIMO, their biological relevance suggests that TIMO provides biological valid estimates of cellular growth dynamics during this stage. Notably, Moscot incorporates prior biological knowledge when estimating growth rates, whereas TIMO does not rely on  prior information. Despite this, TIMO achieves comparable overall performance and improved marker recovery during this critical developmental interval. The distortion metric comparison together with the association analysis using curated gene sets both indicate that TIMO yields more biologically meaningful and reliable estimates of cellular growth rates.

\section{Theoretical Analysis}
\label{sec:theory}
The empirical results in Section \ref{sec:data} demonstrate that TIMO provides biologically meaningful transition estimates across multiple time points. In this section, we study the theoretical properties of the semi-relaxed OT. In particular,  we study the non-asymptotic bound, the asymptotic distribution and the bootstrap consistency of estimated transport matrix and estimated growth rates given by TIMO  under an idealized metacell model.

We assume the oracle metacell construction that the number of metacells and the label of metacells are correctly identified. Let $\Delta^{K-1} = \{\boldsymbol{p}=(p_1,\ldots,p_K): \sum_{k=1}^{K}p_k =1,~p_k> 0~\text{for all}~k\}$ be the probability simplex.
For each time point  $t$,  let $\boldsymbol{p_t}\in \Delta^{m_t-1} $   denote  true metacell proportion, and  let $\boldsymbol{\hat{p}_t}\in \Delta^{m_t-1}$   denote the empirical proportions from multinomial distribution of sizes $N_t$. 
Let $\hat{\mu}_{t,k}$ be the sample centroid of metacell $k$. For notational simplicity,  we focus on two time points $t=0$ and $t=1$ in the theoretical study as the

\vspace{-0.5cm}
\noindent general cases can be derived using identical techniques.   Let $C_{k\ell} = \|\mu_{0,k} - \mu_{1,\ell} \|_2^2$ denote the unknown ground truth of cost matrix, $\widehat C_{k\ell} = \|\hat\mu_{0,k} - \hat\mu_{1,\ell} \|_2^2$ its empirical counterpart. For fixed hyperparameters $\lambda_0>0$ and $\lambda_1>0$, let $\boldsymbol{\pi^*}$ be the population minimizer:

\begin{equation}\label{gt}
\begin{aligned}
\boldsymbol{\pi^*}
= &\arg\min_{\boldsymbol{\pi} \in \mathbb{R}_+^{m_0\times m_{1}}}
\Bigg\{
 \sum_{i=1}^{m_0}\sum_{j=1}^{m_1} \pi_{k\ell}\, \|\mu_{0,k}- \mu_{1,\ell}\|_2^2 
+ \lambda_0 \sum_{k=1}^{m_0}\sum_{\ell=1}^{m_{1}}\pi_{k\ell}
\left(\log {\pi_{k\ell}}-1\right)
\\
&\quad + \lambda_1 \sum_{k=1}^{m_0} \left(\sum_{\ell=1}^{m_{1}} \pi_{k\ell}\right)\log\!\left(\frac{\sum_{\ell=1}^{m_{1}}\pi_{k\ell}}{{p}_{0,k} }\right)
\Bigg\},
\\&~\text{s.t.} \sum_{k=1}^{m_{0}} \pi_{k\ell} = {p}_{1,\ell},~~ \pi_{k,\ell}\geq 0. 
\end{aligned}
\end{equation}
Let $\boldsymbol{\hat{\pi}}$ denote the plug-in estimator of $\boldsymbol{\pi^*}$ in Eq.\eqref{gt}, which corresponds to the estimator in Eq.\eqref{SOT} with $t=0$.  
We impose the following  assumptions:
\begin{enumerate}[label=(\textbf{A\arabic*})]
\item \label{DGM}
$S_{t,i}\sim \text{Multi}(\boldsymbol{p_t}),~X_{t,i}|S_{t,i}=k \sim F_k$, where the distribution $F_{t,k}$ has expectation  $\mu_{t,k}$ and non-singular covariance matrix $\Sigma_{t,k}$ for $k=1,\ldots,m_t$.
\item \label{bounded} There exists a finite constant $M$, such that $\|X_{t,i}\|_2\leq M$ almost surely. 
\end{enumerate}
The  assumption \ref{DGM}  states that for each time point $t$, the underlying biological states can be represented by a finite set of latent states, and the observed data are noise contaminated version of these biological states.  This assumption should be interpreted as a statistical and computationally-convenient representation rather than a claim about biological discreteness. Under finite sequencing depth and technical noise, the data-generating mechanism is only identifiable up to a finite resolution. Even if the underlying biological process evolves continuously, states that are closer than the noise-induced indistinguishability scale cannot be reliably distinguished from the observed data. Thus, representing the system using a finite number of states provides a reasonable statistical approximation. From an approximation-theoretic perspective, for any Gaussian noise-contaminated continuous distribution with $N$ observations, one can construct an $O(\log N)$  mixture model  that is statistically indistinguishable under suitable distributional metrics \citep{polyanskiy2020}, thereby justifying the use of a finite state assumption. 
% The assumption of a diagonal covariance matrix is made for technical convenience in the analysis and can be relaxed without significant additional difficulty.
Let $\text{Vec}(\cdot)$ be the vectorization operator of matrices. We then study both the finite sample bound    and the  asymptotic limiting distribution of $(\text{Vec}(\boldsymbol{\hat{\pi}})-\text{Vec}({\boldsymbol{\pi^*}})) $.  
\begin{theorem}\label{thm1}
Let $N = N_0 + N_1$. Suppose Assumptions \ref{DGM} and \ref{bounded} hold and there exists a constant $\omega \in (0,1)$ such that $\lim_{N \to \infty} \frac{N_0}{N} = \omega $.
Then $\boldsymbol{\hat{\pi}}$ converges to $\boldsymbol{\pi^*}$  
\[
\mathbb{E}\big\|\text{Vec}(\boldsymbol{\hat{\pi}})-\text{Vec}({\boldsymbol{\pi^*}})\big\|_1 \lesssim N^{-1/2},
\]
and
\[
\sqrt{N}\big\{\text{Vec}(\boldsymbol{\hat{\pi}})-\text{Vec}({\boldsymbol{\pi^*}})\big\}
\overset{D}{\rightarrow} \mathcal{N}(0,\Sigma),
\]
where $\Sigma$ is a positive-semidefinite matrix and $\mathcal{N}(0,\Sigma)$ denotes the multivariate normal distribution with mean zero and covariance matrix $\Sigma$.
\end{theorem}

The technical proof of the above theorem is provided in Section S2 of the Supplementary Materials. The theorem shows that when the number of metacells is fixed and the metacell assignments are correctly identified, the estimator  $\boldsymbol{\hat{\pi}}$ is a $\sqrt{N}$-consistent estimate of $\boldsymbol{\pi^*}$. The asymptotic covariance matrix depends on the sampling ratio $\omega$, the unknown population parameters $\boldsymbol{p_0},~\boldsymbol{p_1}$, the expectation and the covariance matrix of $F_k$  and the regularization parameters $\lambda_0$ and $\lambda_1$. Since the explicit form  of the covariance matrix is complicated, we leave it in Section S2of the Supplementary Materials. 
Let $g_k$ denote the relative growth rate of the $k$-th metacell, defined as
$
g_k := \frac{\sum_{\ell} \pi^*_{k\ell}}{p_{0,k}}$.
, whose  corresponding estimator is given by 
$
\hat g_k = \frac{\sum_{\ell} \hat{\pi}_{k\ell}}{\hat p_{0,k}}.$
By   the delta method \citep{van2000asymptotic}, for any $k = 1, \ldots, m_0$, $\hat g_k$ is also a $\sqrt{N}$-consistent estimator of $g_k$, and
$\sqrt{N}(\hat g_k - g_k)$ converges to some normal distributions. 

  We further  have the following theorem regarding the asymptotic distribution of $\boldsymbol{\hat \pi}$. Let $\{(X^b_{t,i},S^b_{t,i})\}_{i=1}^{N_t}$ be the empirical bootstrap replicate of  $\{(X_{t,i},S_{t,i})\}_{i=1}^{N_t}$, and $\boldsymbol{\tilde{\pi}}$ be the estimate derived from the bootstrap replicate of  $\{(X_{0,i},S_{0,i})\}_{i=1}^{N_0}$ and $\{(X_{1,j},S_{1,j})\}_{i=1}^{N_1}$. The bootstrap consistency is given as follows.
\begin{theorem}\label{cor1}Suppose the conditions in Theorem \ref{thm1} hold. Let $\boldsymbol{Z}$ a random variable drawn from the asymptotic distribution of  $\boldsymbol{\widehat{\pi}}-\boldsymbol{{\pi}^*}$  derived from  Theorem \ref{thm1}.  
The empirical bootstrap procedure is consistent in the sense that 
$$\sup_{f\in BL_1(\mathbb{R}^{m_0\times m_1})} \bigg|\mathbb{E}\big[ f\big(\sqrt{N}\{\boldsymbol{\tilde{\pi}}-\boldsymbol{\widehat{\pi}}\} \big) |\{X_{0,i}\},\{X_{1,j}\} \big]- \mathbb{E} f(\boldsymbol{Z})\bigg|\to 0~\text{in probability} ,$$
where $BL_1(\mathbb{R}^{m_0\times m_1})$ is the function space defined on $\mathbb{R}^{m_0\times m_1}$  consisting of all functions that are uniformly Lipschitz with Lipschitz constant $1$ w.r.t Euclidean distance. 
\end{theorem} 

 Our theoretical results differ from existing asymptotic analyses of entropic optimal transport. Prior work typically focuses on the balanced OT setting, where both marginal constraints are enforced \citep{klatt2020empirical}. In contrast, our analysis considers a semi-relaxed OT formulation, where only one marginal constraint is imposed. A second key distinction is that the cost matrix in our framework is random and estimated from the data, whereas   existing results assume the transport cost matrix is known a priori. To the best of our knowledge, this is the first work to establish asymptotic convergence results for semi-relaxed optimal transport under a finite mixture model with data-driven (random) cost matrices. In Section S3, we conduct simulation studies to empirically validate the convergence of $\boldsymbol{\hat{\pi}}$ to $\boldsymbol{\pi}^*$ under both total variation and KL divergence, across a range of metacell numbers (mixture components) and graining levels (average number of cells per metacell) as $N\to\infty$. We also  perform simulation study to numerically support bootstrap consistency in Section S3 of the Supplementary Materials.
 % \kong{It might be a bit better to move this part to the analysis section.}
The theoretical analysis above relies on an idealized metacell construction, where both the number of metacells and the metacell assignments are assumed to be correctly identified.  In practice, metacells are typically constructed through a combination of dimensionality reduction and assignment procedures, and different implementations may lead to different metacell. Providing a rigorous theoretical analysis that jointly accounts for the uncertainty introduced by these unsupervised learning steps and the subsequent optimal transport estimation remains a challenging problem. In particular, the statistical theory of clustering and other unsupervised learning methods in high-dimensional settings is still an open problem and an active area of research. Though the theoretical justification relies on idealized  metacell construction,   we show in Section S4 that the limiting distribution of the TIMO estimator is well approximated by both the bootstrap and normal distributions in our real data case. Developing a unified theoretical framework for metacell  and optimal transport  is still  an interesting  direction for future work.

\section{Discussion}\label{sec:diss}
In this paper, we introduce TIMO, a framework for studying the time course of  scRNA-seq developmental data. TIMO is a two-stage procedure that combines the ideas of metacells and optimal transport. In the first stage, cells are aggregated into homogeneous groups, referred to as metacells, such that cells within the same metacell are assumed to share the same underlying biological state, and variations within each group are treated as technical noise.    The second stage of TIMO therefore estimates the transport matrix between metacells at two consecutive time points by solving a simpler semi-relaxed optimal transport problem.

We apply TIMO to a large-scale mouse developmental atlas dataset and benchmark its performance against  Moscot in estimating cell-type transition probabilities and cellular growth rates. Our results show that TIMO provides  biologically valid estimates for both quantities. Although TIMO is conceptually simple, it significantly reduces the computational burden of solving optimal transport problems and scales well to large datasets, particularly when millions of cells are sequenced.  In addition, we establish theoretical guarantees for TIMO, including non-asymptotic error bounds, asymptotic distributions and the bootstrap consistency of the estimators for confidence interval construction.  To the best of our knowledge, this is the first statistical analysis of semi-relaxed optimal transport under a setting where the transport cost matrix is estimated from the data and the first attempt to perform uncertainty quantification for OT-based single-cell trajectory inference.

However, there are still several limitations of TIMO. First, it is generally not possible to uniquely disentangle changes in cellular distributions into growth and differentiation components from snapshot single-cell data. This identifiability issue is not specific to our method but represents a broader challenge in trajectory inference from snapshot observations. 
 Second, our theoretical analysis relies on an idealized assumption that the metacell construction is  oracle, meaning that the number of metacells and the cell assignments are correctly identified. In practice, metacells are obtained through a combination of dimensionality reduction and  assignment procedures. Developing statistical theory that jointly accounts for these unsupervised learning steps together with optimal transport estimation remains an  open problem. Another promising direction is to extend the TIMO framework to multiomics data, where multiple modalities such as spatial locations,  gene expression, chromatin accessibility, and protein abundance are jointly measured, allowing for a more comprehensive characterization of developmental dynamics.
 We leave these problems as our future research.  

 % \kong{we could mention what are included in the supplementary materials, say simulations, additional data applications, etc.}

% $$S_{t,i}\sim \text{Multi}(p_t),~\mathbb{E}[X_{t,i}|S_{t,i}=k] = \mu_{t,k},~\mathbb{E}[(X_{t,i}-\mu_{t,k})(X_{t,i}-\mu_{t,k})^\top |S_{t,i}=k]=\sigma^2 I. $$

\bibliographystyle{asa}

\bibliography{ref}

\end{document}